\documentclass[10pt]{article}
\usepackage{amssymb,citesort}
\usepackage{multicol,epsfig}
\begin{document}

\title{Absence of the liquid phase when the attraction is not
pairwise additive}

\author{{\bf Richard P. Sear}\\
Department of Physics,
University of Surrey, Guildford, Surrey GU2 5XH, United Kingdom}


\maketitle

\begin{abstract}
Recent work on charged colloidal suspensions with very low levels
of added salt has suggested that although pairs of the colloidal
particles repel, clusters of the particles attract.
Motivated by this, we study simple model particles which have
many-body attractions. These attractions are generic
many-body attractions and are not calculated for any specific
colloidal suspension. We find that many-body
attractions can stabilise solid phases at low pressures but that
the liquid phase is either completely absent from the equilibrium
phase diagram or present only within a small region of parameter space.
\end{abstract}

\begin{multicols}{2}

\section{Introduction}

Although the behaviour of suspensions of highly charged colloidal
particles at very low added salt concentrations is highly controversial,
accurate measurements on isolated pairs of the particles shows nothing
but a pure repulsion \cite{crocker96}
yet the particles form crystallites which appear
to be metastable at close to zero osmotic pressure \cite{ito94}.
This is puzzling, particles interacting via a pairwise additive,
purely repulsive interaction should only form a solid phase at
high osmotic pressure, the spheres need to be pushed together against
their repulsion. The solid phase of hard spheres is certainly
not metastable at low pressure.
Thus the observation of the repulsion of isolated
pairs, and of the clustering of larger numbers of spheres is
inconsistent with a description of the interaction of the colloidal
particles via a pairwise additive potential of mean force.
If we are to continue to describe these systems as particles
interacting via some potential of mean force, then we must relax
our restriction to a pairwise additive potential of mean force
and consider a many body potential \cite{note1}. We do this here. We consider
simple, rather generic, many-body attractions and calculate
the phase behaviour they lead too. We find that if pairs of particles
repel but triplets attract that the phase behaviour is qualitatively
similar to that found for a conventional pairwise additive attraction
except that the region within which the liquid phase is found is much
smaller. If pairs and triplets repel but four particles attract then
there is no liquid phase at all; the attraction merely broadens
the coexistence region between the fluid and solid phases.
The experiments find metastable solid phases but no liquid phases,
except for one, controversial, result \cite{tata92}.

Here, we do not address the question of what is
the origin of a purely repulsive interaction between pairs of
particles but an attraction between larger numbers of particles.
We do {\it not} attempt to derive the many-body attraction
from the electrostatic interactions
between the charged colloids and the counter and coions.
However, there has been some work in which
the existence, or non-existence, and origin of many-body
interactions in suspensions of highly charged colloids has been
considered \cite{schmitz,ha}. Also, van Roij, Dijkstra and Hansen \cite{roij99}
have found both fluid-solid and fluid-fluid coexistence
within an approximate theory.
We are simply interested in the consequences, particularly the
phase behaviour, of many-body attractions. Because we have not
derived our interaction potentials from the
underlying electrostatic interactions we cannot say anything
quantitative about the suspensions of highly charged colloidal
particles. However, our results show that rather generally
if pairs repel then a many-body attraction can still cause
condensed phases to appear but that these condensed phases are
much more likely to be solid rather than liquid phases.
This conclusion only relies on the attractions being many-body, it
does not rely on, for example, whether or not the particles are charged.

In order to keep our model of non-pairwise additive attractions
as simple as possible we simply generalise a simple van der Waals-type
model. This model is that of particles interacting via a hard-sphere
repulsion and a long-range pair attraction. It has been extensively
studied, see Refs. \cite{vdw,hansen86,kampen64,lebowitz66,longuet64}.
We simply generalise the long-range pairwise additive attraction
to a long-range attraction between $n$ particles, i.e., an
$n$-body attraction. This is done in section 2. In section 3 we apply
the standard perturbation theory to obtain the contribution
of the attractions to the free energy. Then in section 4 we
show and discuss the phase behaviour before ending with
a conclusion, section 5.

\section{Potential}

Our potential consists of a hard-core interaction which is pairwise additive
--- the hard-sphere potential --- plus an $n$-body attraction.
The energy of $N$ particles is
\begin{equation}
V({\bf r}^N,\epsilon_n)=V_{hs}({\bf r}^N)+V_a({\bf r}^N,\epsilon_n),
\label{potfunc}
\end{equation}
where ${\bf r}^N$ denotes the $N$ position vectors of the $N$
spherical particles, and $\epsilon_n$ is a measure of the strength
of the $n$-body attraction.
It is the sum of two terms.
$V_{hs}$ is the potential energy of $N$ hard spheres, given by
\begin{equation}
V_{hs}({\bf r}^N)=\frac{1}{2}\sum_{i=1,N}\sum_{j=1,N}^{'}
\phi_{hs}(|{\bf r}_i-{\bf r}_j|)
\end{equation}
where $\phi_{hs}$ is the hard-sphere pair potential,
\begin{equation}
\phi_{hs}(r)=
\left\{
\begin{array}{ll}
\infty & ~~~~~~ r \le \sigma\\
0 & ~~~~~~ r > \sigma\\
\end{array}\right. ,
\label{hspot}
\end{equation}
$\sigma$ is the hard-sphere diameter.
$V_a$ is the energy of attraction coming from an $n$-body
attractive potential $\phi_a$,
\begin{equation}
V_a({\bf r}^N)=
\frac{1}{n!}\sum_{i=1,N}\cdots\sum_{l=1,N}^{'}
\phi_a\left(\left\{ |{\bf r}_j-{\bf r}_k| \right\}^n\right).
\label{vadef}
\end{equation}
The $n$-fold sum is over the $N!/(N-n)!$ sets of $n$ out of the $N$
particles formed by choosing one of the $N$ particles to be particle
$i$, then one of the remaining $N-1$ particles to be particle $j$,
$\ldots$ and then one of the $N-n+1$ particles as particle $l$.
The dash on the sum over $l$ indicates that the sum excludes
terms in which $l=i,j,\ldots$.
This overcounts by a factor of $n!$ as it counts each interaction
$n!$ times, each time with the $n$ particles in a different order.
For example, for $n=2$ it counts the interaction between particles
numbers 11 and 103, for example, twice, once as $i=11$, $j=103$ and once
as $i=103$, $j=11$. For general $n$ the number of ways a set of $n$
particles can be assigned to the $n$ indices $i$, $j$, $\ldots$, $l$ is
$n!$. The $n$-body attractive potential $\phi_a$ is a non-positive function
of the set of $n(n-1)/2$ pair separations of the $n$-particles.
This set of scalar separations is denoted by
$\left\{ |{\bf r}_j-{\bf r}_k| \right\}^n$.
The potential $\phi_a$
is symmetric with respect to exchanging any pair of the $n$ particles.
We write it as the product of a strength of attraction $\epsilon_n$,
and a function $\zeta(\left\{|{\bf r}_j-{\bf r}_k|\right\}^n)$
which determines the dependence of the energy on the particle
coordinates
\begin{equation}
\phi_a(\left\{|{\bf r}_j-{\bf r}_k|\right\}^n)=-\epsilon_n
\zeta(\left\{|{\bf r}_j-{\bf r}_k|\right\}^n),
\label{mbpotdef}
\end{equation}
where $\zeta$ is a non-negative function of the set of
pair separations of the $n$ particles which we need not specify
explicitly but which must be long
ranged, i.e, it decays to zero over some characteristic range much larger than
the hard-sphere diameter $\sigma$. For a configuration in which
all $n(n-1)/2$ separations $|{\bf r}_j-{\bf r}_k|$ are not
much more than the hard-sphere diameter,
$\zeta(\left\{|{\bf r}_j-{\bf r}_k|\right\}^n)\lesssim1$. Then
if one or more of the separations is increased
$\zeta$ tends to zero over a range which is much larger than the
hard-sphere diameter.
The integral of $\zeta$ over the positions of all $n$-particles
is proportional to the volume $V$ in the thermodynamic limit
\begin{equation}
\int\zeta\left(
\left\{|{\bf r}_j-{\bf r}_k|\right\}^n\right){\rm d}{\bf r}^n=V\nu_n.
\label{zetaint}
\end{equation}
$\nu_n$ is finite and a constant; it
has dimensions of length to the power of $3(n-1)$.

\section{Thermodynamic functions}

As the attraction is assumed to be long-ranged, i.e., with a range
much larger than the hard-sphere diameter $\sigma$, we may use the van der
Waals approximation of approximating the free energy by the
free energy of a system interacting only via the repulsive part of
the potential plus the energy of interaction due to the attractive
part of the potential. Here the system interacting only via
the repulsions is that of hard spheres, whose free energy in the
fluid and solid phases are accurately given by the approximations
of Carnahan and Starling \cite{carnahan69},
and of Hall \cite{hall72}, respectively.
So, we approximate the Helmholtz free energy by that of hard spheres
$A_{hs}$, plus
the energy (not the free energy) evaluated by assuming that the
configurations of the particles are unaffected by the attraction, $U$
\begin{equation}
A(N,V,T)=A_{hs}(N,V,T)+U(N,V).
\label{avdw}
\end{equation}
$A_{hs}$ is given by
\begin{equation}
A_{hs}=-kT\ln Z_{hs}\Lambda^N
\end{equation}
where
\begin{equation}
Z_{hs}=\frac{1}{N!}\int\exp\left[-\beta V_{hs}({\bf r}^N)\right]{\rm d}{\bf r}^N
\end{equation}
is the configuration integral for $N$ hard spheres in a volume $V$.
$\Lambda$ is the thermal volume of a particle.
$U$ is given by
\begin{equation}
U=\frac{\int
V_a({\bf r}^N)
\exp\left[-\beta V_{hs}({\bf r}^N)\right]
{\rm d}{\bf r}^N}{Z_{hs}}
\label{udef}
\end{equation}

If we substitute our expression for $V_a$, Eq. (\ref{vadef}),
into Eq. (\ref{udef}) we note
that it is the sum of $N!/(N-n)!$ equivalent terms \cite{note}. So,
\begin{eqnarray}
U&=&\frac{N!}{(N-n)!n!}\times\nonumber\\&& \frac{\int
\phi_a\left(\left\{ |{\bf r}_j-{\bf r}_k| \right\}^n\right)
\exp\left[-\beta V_{hs}({\bf r}^N)\right]
{\rm d}{\bf r}^N}{Z_{hs}},
\label{udef2}
\end{eqnarray}
but the $n$-particle density of hard spheres, $\rho^{(n)}_{hs}$,
and their $n$-particle distribution function, $g^{(n)}_{hs}$,
are defined as \cite{hansen86}
\begin{eqnarray}
\rho^{(n)}_{hs}({\bf r}^n)&=&
\left(\prod_{i=1,n}\rho^{(1)}_{hs}
({\bf r}_i)\right) g^{(n)}_{hs}({\bf r}^n)\nonumber\\
&=&\frac{N!}{(N-n)!}\frac{\int
\exp\left[-\beta V_{hs}({\bf r}^N)\right]
{\rm d}{\bf r}^{N-n}}{Z_{hs}}.
\label{gdef}
\end{eqnarray}
The 1-particle density, $\rho^{(1)}_{hs}({\bf r})$,
is not assumed to be uniform
so that the theory applies to solid as well as fluid phases.
Using, Eq. (\ref{gdef}) in Eq. (\ref{udef2}), we obtain
\begin{equation}
U=\frac{1}{n!}\int
\phi_a\left(\left\{ |{\bf r}_j-{\bf r}_k| \right\}^n\right)
\left(\prod_{i=1,n}\rho^{(1)}_{hs}({\bf r}_i)\right)
g^{(n)}_{hs}({\bf r}^n){\rm d}{\bf r}^n.
\label{udef3}
\end{equation}
We now use the long-range of the potential $\phi_a$ to simplify
Eq. (\ref{udef3}). As $\phi_a$ decays to zero only
when the separations of the particles are much larger than $\sigma$,
the integral for
$U$ is dominated by configurations when the spheres are far apart, i.e.,
when all $n(n-1)/2$ pair separations are much larger than $\sigma$.
In a fluid the one particle density is a constant, $\rho^{(1)}=\rho$,
and at separations large with respect to $\sigma$ the distribution
function is close to one, $g^{(n)}_{hs}\simeq 1$. Thus in the fluid phase
the integrand of Eq. (\ref{udef3}) is simply approximated by
$\rho^n\phi_a$.
In a solid phase, although there are long-range
correlations in $\rho^{(n)}_{hs}$ the one particle
density, $\rho^{(1)}_{hs}$, averaged over a unit cell is
just $\rho$.
The attractive interaction between particles has a range much larger than
than the lattice constant of the lattice and so $\phi_a$ varies
little across a unit cell and we can regard $\rho^{(1)}_{hs}({\bf r})$
as approximately constant at its average value, $\rho$.
Similarly, as we change any one of the $n$ position vectors
upon which the $n$-body distribution function, $\rho^{(n)}_{hs}$, depends
the density oscillates rapidly over each unit cell but averages to
$\rho$. So, we approximate $\rho^{(n)}_{hs}$ by $\rho^n$.
Then the integrand of Eq. (\ref{udef3}) is the same as in a fluid phase,
$\rho^n\phi_a$.
So, we approximate $g^{(n)}_{hs}$ by one, and
$\rho^{(1)}_{hs}$ by $\rho$ in Eq. (\ref{udef3}) for both fluid
and solid phases,
\begin{equation}
U=\frac{\rho^n}{n!}\int
\phi_a\left(\left\{ |{\bf r}_j-{\bf r}_k \right\}^n\right)
{\rm d}{\bf r}^n,
\end{equation}
then using Eqs. (\ref{mbpotdef}) and (\ref{zetaint}) we obtain
our final expression for the energy
\begin{equation}
U=-\frac{N\rho^{n-1}\epsilon_n \nu_n}{n!}=-N\alpha_n\rho_r^{n-1}kT,
\label{ufinal}
\end{equation}
where $\rho_r=\rho\sigma^3$ is a reduced density and
\begin{equation}
\alpha_n=\frac{\epsilon_n\nu_n}{n!\sigma^{3(n-1)}kT},
\end{equation}
is inversely proportional to the temperature.

Using, our expression for the energy, Eq. (\ref{ufinal}) in the
van der Waals approximation for the Helmholtz free energy, Eq. (\ref{avdw})
yields
\begin{equation}
\beta a(\rho,T)=\beta a_{hs}(\rho)-\alpha_n\rho_r^{n-1},
\label{amb}
\end{equation}
where $a=A/N$ and $a_{hs}=A_{hs}/N$. The pressure
times $\sigma^3$, $p$, is easily
obtained by differentiating the free energy, Eq. (\ref{amb}),
\begin{equation}
\beta p(\rho,T)= \beta p_{hs}(\rho) -(n-1)\alpha_n\rho_r^n.
\label{pmb}
\end{equation}
$p_{hs}$ is the pressure of hard spheres times $\sigma^3$.
The chemical potential $\mu$ is then obtained from
$\mu=a+p/\rho_r$. Knowledge of the pressure and chemical potential is
enough to determine the phase diagram.

For $n=2$, Eq. (\ref{amb}) reduces to a corrected
version of the free energy derived by van der Waals
120 years ago \cite{vdw,hansen86,lebowitz66}.
By corrected we mean that van der Waals' crude approximation
for the free energy of particles interacting via a strong repulsion has
been replaced by the free energy of hard spheres.
Van Kampen \cite{kampen64}, and
Lebowitz and Penrose \cite{lebowitz66} were able to show that
if $a_{hs}$ is exact then Eq. (\ref{amb}) for $n=2$ is exact in the
limit that the range of the attractive potential tends to infinity.
For finite range it is a reliable and very widely used approximation.

\section{Phase behaviour}

Hard spheres freeze at high densities, i.e., they undergo
a first-order transition to the solid phase \cite{hansen86}.
The transition occurs between a fluid phase at a volume
fraction $\eta=(\pi/6)\rho\sigma^3=0.49$ and a solid phase
at the higher volume fraction $\eta=0.55$
\cite{wood58,alder58,hoover68}. This is the only transition
in the $\alpha_n=0$ limit. The $\alpha_n$ are integrated
strengths of attraction divided by temperature and so
$1/\alpha_n$ is a reduced temperature. Thus, the
$\alpha_n=0$ limit is equivalent to the $T\rightarrow\infty$ limit.
However, as the temperature decreases
$\alpha_n$ increases. The first effect this has is to widen the
fluid-solid coexistence region due to the fact that the denser
solid phase has its chemical potential and pressure lowered more
than the less dense fluid.
But at sufficiently low temperature the pressure in a single phase
can become a non-monotonic function of density
--- the negative term due to attractions in the
pressure, Eq. (\ref{pmb}) creates a dip in the pressure --- this is
a van der Waals loop and it indicates that there is phase separation
into dilute and dense phases of the same symmetry.
The signature of phase coexistence is a van-der-Waals loop as
our theory is a mean-field theory; it will therefore predict
incorrect critical exponents.

\begin{figure}
\caption{
The phase diagram of hard spheres plus a long-range pairwise
additive attraction, $n=2$. The $x$-axis is a reduced density, the
volume fraction $\eta$, and the $y$-axis is a reduced temperature,
$1/\alpha_2$.
The thick solid curves separate the one and two-phase regions.
The letters V, L and S denote the regions of the phase space
occupied by the vapour, liquid and solid phases, respectively.
The horizontal thin lines are tie lines connecting coexisting densities.
}
\label{fig1}
\begin{center}
\epsfig{file=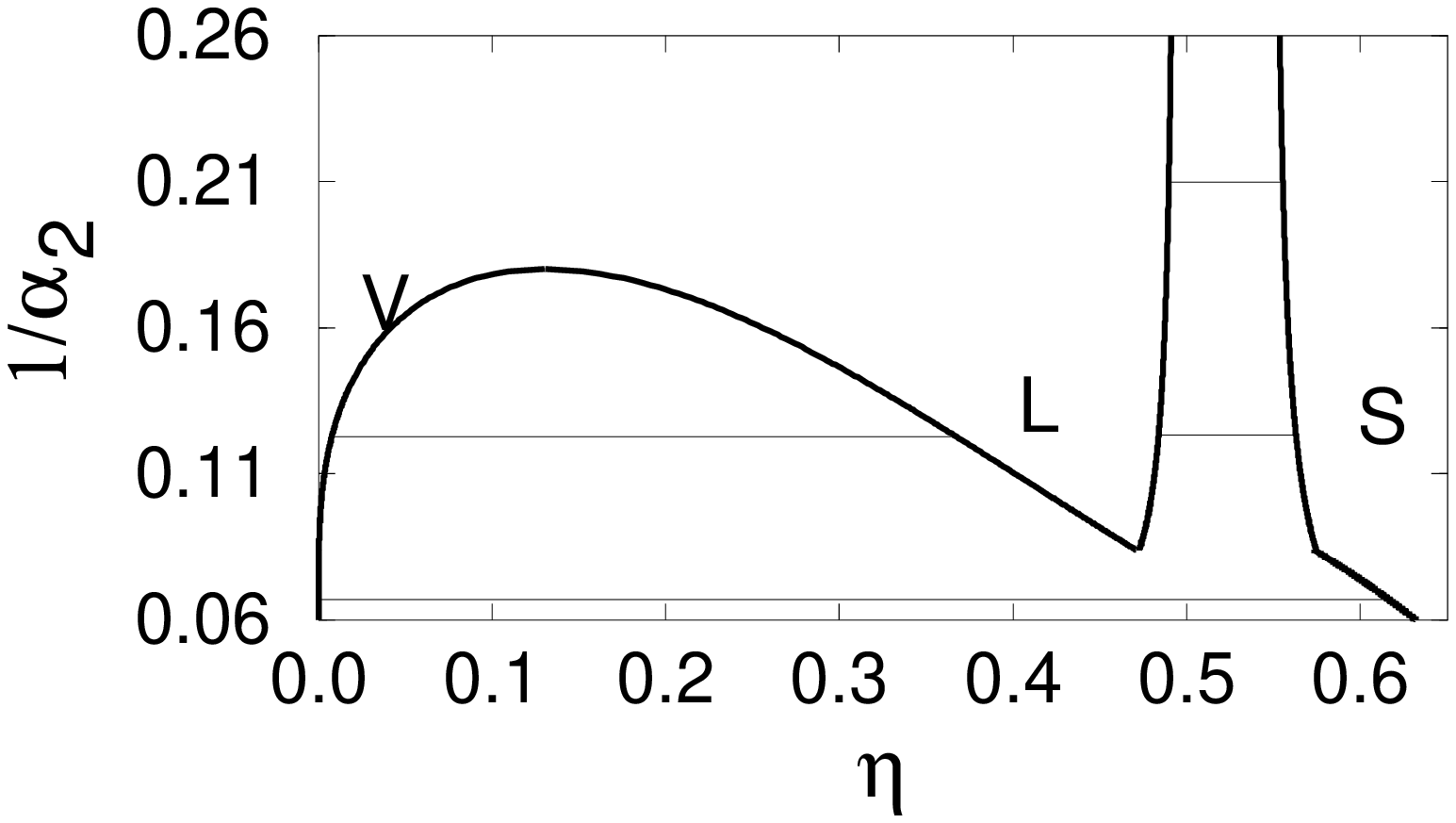,width=4.0in}
\end{center}
\end{figure}

For $n=2$, we have particles interacting via a hard-sphere repulsion
and a long-range pairwise additive attraction: essentially the model
postulated by van der Waals in the last century to describe the vapour-liquid
transition. It, of course, has a vapour-liquid transition, as can be seen
in Fig. \ref{fig1} which shows its phase behaviour in the
density-temperature plane. There is a large temperature range over which
there are coexisting dilute (vapour) and dense (liquid) fluid phases:
the temperature at the critical point is more than twice its value
at the triple point. The critical point is the maximum in the
vapour-liquid coexistence curve and occurs at $1/\alpha_2=0.18$,
and the triple point temperature is the temperature at which vapour,
liquid and solid phases coexist. It marks the lower limit
of equilibrium vapour-liquid coexistence and occurs at
$1/\alpha_2=0.084$.

\begin{figure}
\caption{
The phase diagram of hard spheres plus a long-range three-body
attraction, $n=3$.
See caption of Fig. \ref{fig1} for details.
}
\label{fig2}
\begin{center}
\epsfig{file=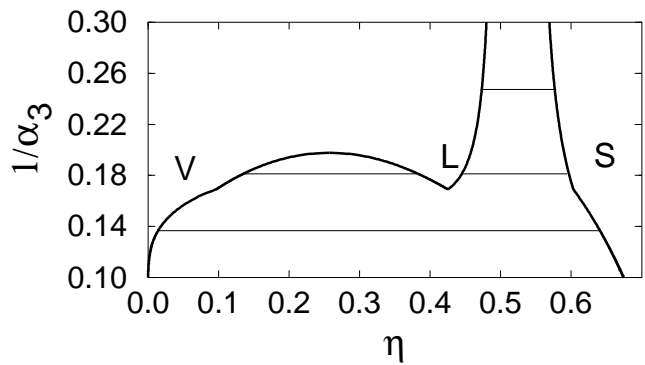,width=4.0in}
\end{center}
\end{figure}

For $n=3$, although there is still equilibrium vapour-liquid
coexistence, see Fig. \ref{fig2}, the temperature range over which
it occurs is much reduced. At the critical and triple point
temperatures, $1/\alpha_3=0.20$ and 0.17, respectively. The ratio
between the two temperatures is only 1.2. The density at the
critical point is also considerably higher than for a pairwise
additive attraction. 

\begin{figure}
\caption{
The phase diagram of hard spheres plus a long-range four-body
attraction.
The letters F and S denote the regions of the phase space
occupied by the fluid and solid phases.
The dashed curve is the coexistence curve for a
vapour-liquid transition within the fluid-solid coexistence region.
See caption of Fig. \ref{fig1} for further details.
}
\label{fig3}
\begin{center}
\epsfig{file=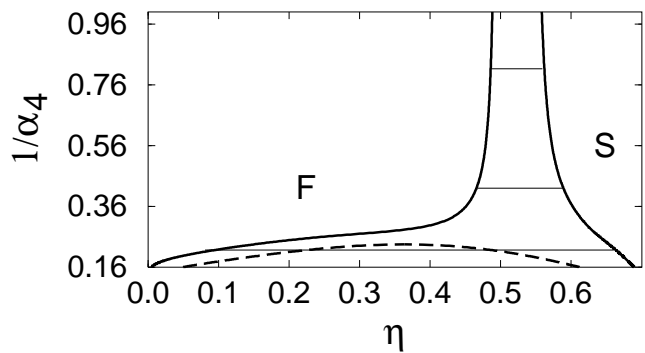,width=4.0in}
\end{center}
\end{figure}

For $n=4$ there is no equilibrium vapour-liquid coexistence,
see Fig. \ref{fig3}.
At equilibrium there is only one phase transition: the fluid-solid transition.
The vapour-liquid transition has not disappeared without trace, however.
Our mean-field pressure in the fluid phase does develop a van der
Waals loop at low temperature and we can construct a vapour-liquid
coexistence curve {\it within} the fluid-solid coexistence region.
This is plotted as a dashed curve in Fig. \ref{fig3}. Note that
the critical density is even higher than for a 3-body attraction.
The disappearance of the equilibrium vapour-liquid coexistence
is just what is observed when the range of a pairwise additive
attraction is reduced so that it is only about 10\% or less of
the hard-sphere diameter,
see Refs.
\cite{gast83,lekkerkerker92,hagen94,bolhuis94,daanoun94,dijkstra99}.

For $n$ larger than 4, the vapour-liquid transition moves deeper
into the fluid-solid coexistence region, the density at the
critical point increases with increasing $n$. If $n$ is made
extremely large then a van der Waals loop appears in the
pressure of the {\it solid} phase. By $n=15$ there is
equilibrium coexistence between two solid phases of the same
symmetry but different densities. This is just as has been found with
a very short-range attraction, a few \% of the hard-sphere diameter
\cite{bolhuis94}.

The physical interpretation of the phase behaviour is straightforward. The
energy per unit volume
due to an $n$-body attraction is proportional to the number
of clusters of $n$ particles where all $n$ particles are close
enough to each other to be within range of the attraction. It
is this dependence which leads directly to the $\rho^n$ dependence
of the energy per unit volume. So, the larger $n$ is, the smaller
the number of clusters of $n$ particles at low density and the more rapidly
the number of clusters grows as density increases.
When $n$ is small, $n=2$, the energy starts to become significant at
a relatively low
density and it decreases gently, as density squared. This creates a
van der Waals loop in the pressure at low density. The slow
decrease in energy as density increases allows the hard-sphere
part of the free energy to dominate at densities which are not too high.
This means that the pressure turns up at densities below freezing,
so allowing a stable liquid to coexist with the vapour.
For $n=2$ the van der Waals loop is
below the density at which hard-spheres solidify and we see equilibrium
fluid-fluid coexistence. However, for $n>2$, the energy only
becomes significant at higher densities and it decreases rapidly
with increasing density.
The more rapid decrease with density of many-body
attractions directly favours the solid phase simply because it is
more dense.  This then broadens the fluid-solid coexistence
region to such an extent that the fluid-fluid transition becomes
metastable. Also, as the energy only decreases at relatively high
density for large $n$, the
van der Waals loop is shifted to higher density; ultimately
for $n=6$ the critical point is at such high density,
a volume fraction $\eta=0.51$, that it is above the density
at which the particles freeze. It is therefore not observable.

To summarise: a combination of the increased
stability of the denser solid phase with respect to the less dense
fluid phase, and the increasing density of critical point drive
the fluid-fluid transition metastable, then, as we continue
to increase $n$, the critical density becomes so large that it
exceeds the density at which hard spheres crystallise. Beyond
this point it is not possible to observe fluid-fluid coexistence.

\section{Conclusion}

We have shown that many-body
attractions have a much weaker tendency to induce separation of
the fluid phase into dilute (vapour) and dense (liquid) phases than
do pairwise additive attractions. Indeed when not only pairs
but triplets of particles interact via a pure repulsion and with only
four or more particles is there attraction, we found no equilibrium
coexistence between dilute and dense fluid phases.
Therefore, we expect that the findings for highly charged
colloidal particles are general in the following sense:
if pairs of particles repel then some systems may still have an attraction
between larger numbers of particles. This may cause the
fluid-solid transition to broaden greatly and produce a solid
phase at much lower (osmotic) pressures than in a purely repulsive
system, however, there will either be only a small range of parameters over
which there is equilibrium fluid-fluid coexistence or no
equilibrium fluid-fluid coexistence at all.
Of course there are an infinity of possible $n$-body attractions
and they will lead to different behaviour in the same way that different
forms of pairwise attraction leads to different behaviour. However,
our results should apply to potentials which are reasonably smooth and
long-ranged, i.e., with a range of roughly
the hard-sphere diameter or more. We are also assuming that the
particles are spherical.

Finally, there has been a great deal of interest in the disappearance of the
liquid phase as the potential is varied; attention has focused
on particles in which the range of a pairwise additive
attraction is very short
\cite{gast83,hagen94,bolhuis94,lekkerkerker92,daanoun94,dijkstra99}.
But for other examples of liquid phases disappearing see Ref. \cite{faraday}.
The behaviour we have found as we went from 2 to 3 to 4-body attractions
is qualitatively exactly the same as has been found as
the range of a pairwise additive attraction is made very short.

It is a pleasure to acknowledge discussions with D. Frenkel, B.-Y. Ha and R.
van Roij.


\end{multicols}

\newpage

\vspace{-2.0in}

\begin{thebibliography}{99}
\bibitem{note1} We do not, of course, have to continue to describe
the colloidal suspension in terms of particles interacting
via a potential of mean force. We can revert to treating the
counter and coions explicitly.
\bibitem{note} Our derivation of Eq. (\ref{ufinal}) for the energy
is a straightforward generalisation from 2 to $n$-body potentials
of a textbook derivation, see Ref. \cite{hansen86}.
\end{thebibliography}
\end{document}